\documentstyle[12pt]{article}                          
\catcode`@=11
\def\secteqno{\@addtoreset{equation}{section}%
\def\theequation{\thesection.\arabic{equation}}}
\catcode`@=12
\newcommand{\be}{\begin{equation}}
\newcommand{\ee}{\end{equation}}
\newcommand{\bea}{\begin{eqnarray}}
\newcommand{\eea}{\end{eqnarray}}

\newcommand{\sltp}{/ {\hskip-0.27cm{{\wp}}}}
\newcommand{\sltP}{/ {\hskip-0.27cm{\cal {P}}}}
\newcommand{\slPi}{/ {\hskip-0.27cm{\Pi}}}
\newcommand{\slpi}{/ {\hskip-0.27cm{\pi}}}

\begin{document}
	  \hfill TOHO-FP-9858

	  \hfill YITP-98-4

\vskip 20mm 
\begin{center} 
{\bf \Large Canonical Equivalence between 
Super D-string and Type IIB Superstring} 
\vskip 10mm
{\large Yuji\ Igarashi, Katsumi\ Itoh,$^a$\footnote{On leave of
absence from Faculty of Education, Niigata University, Niigata 950-21,
Japan} Kiyoshi\ Kamimura,$^b$ and Rie\ Kuriki$^c$}\par 

\medskip
{\it 
Faculty of Education, Niigata University, Niigata 950-21, Japan\\
$^a$ Yukawa Institute for Theoretical Physics, Kyoto University\\Kyoto
606-01, Japan\\
$^b$ Department of Physics, Toho University, \ Funabashi\ 274, Japan\\
$^c$ Research Institute of Electrical Communication\\
2-1-1 Katahira, Aoba-ku, Sendai\ 980-8577, Japan\\
}

\medskip
\date{\today}
\end{center}
\vskip 10mm
\begin{abstract}

We show that the super D-string action is canonically equivalent to
the type IIB superstring action with a world-sheet gauge
field. Canonical transformation to the type IIB theory with dynamical
tension is also constructed to establish the $SL(2,Z)$ covariance beyond
the semi-classical approximations.

\end{abstract}
\noindent

\newpage
\setcounter{page}{1}

\parskip=7pt
\section{Introduction}
\indent

Much attention has been focused recently on effective action approach
to D-branes, recognized as a key concept in understanding of string
dualities \cite{Polchinski}.  
The Dirac-Born-Infeld (DBI)
actions have been considered to describe collective motions of the
string solitons in the effective field theory
\cite{Tseytlin},
\cite{Leigh}.
They have been
constructed by the requirement of the conformal
invariance \cite{Leigh}, and their supersymmetric extensions with kappa
symmetry have been given for allowed world volume dimensions
\cite{Cederwall}, \cite{Aganagic},
\cite{Popescu}.\footnote{The explicit form of the Wess Zumino action
for general (IIB) D-branes is given systematically in \cite{Kamimura}.}
Here these extended actions are referred to as D-brane actions.

In this paper we consider the $SL(2,Z)$ symmetry of the type IIB string
theory \cite{Schwarz}.\footnote{In discussing the string duality, one has
to specify the supergravity sector.  As for supergravity backgrounds, we
take the flat metric and include a dilaton and an axion
backgrounds. Both of these are assumed to be constant in order not to
spoil the kappa symmetry.} In the effective action approach, the
central machinery for discussing this symmetry is the ``vector duality"
transformations acting on world-volume gauge fields in the D-brane
actions \cite{Tseytlin}. Several authors have argued that the transformed
D-string (D1-brane) action takes an $SL(2,Z)$ covariant form, and the
D3-brane action is invariant under the vector duality
transformation \cite{Tseytlin}, \cite{Aganagic2}, \cite{Alwis}. These
have been shown first for the bosonic sector, and then extension to the
super D-branes has been given. Although the discussions provide us again
with a support for the S duality, they share a drawback in common being
based on semi-classical approximations. If the S duality is really a
symmetry in string theories, it is natural to expect that one can
establish, beyond semi-classical approximations, the $SL(2,Z)$ covariance
of the D-string theory as well as that of the D3-action.  It is the
purpose of this paper to show that these desired properties of the
D-branes are precisely described via canonical transformations.
The idea that duality transformations can be essentially identified
with canonical transformations was already suggested in \cite{Lozano}
for bosonic truncation of D3-brane and D-string. However the
discussion was only made for the world-volume gauge fields and in the
gauge fixed actions.  We explore the whole structure of the canonical
transformations.  Actually, the canonical transformations we construct
for D-string depend not only on the world-sheet gauge field but also
on the string co-ordinates, especially the fermionic ones.
We report here our results on D-strings only, because we found that
the role of the canonical transformations are different, conceptually
and technically as well, for D-strings and D3-branes.  The D3-action
will be investigated in a subsequent publication \cite{Igarashi}.

Our discussions are based on the canonical Hamiltonian formalism.  The
canonical Hamiltonian of a reparametrization invariant system is
solely expressed by constraints, $\{\varphi_{\alpha}\}$. The Lagrange
multipliers for first-class constraints are undetermined reflecting
underlying gauge symmetries, while those for second-class ones are
fixed by consistency conditions.  The system is thus known to be
purely determined by its constraint surfaces, $\varphi_{\alpha}=0$.
Suppose we have two such systems described by $\{\varphi_{\alpha}\}$
and $\{{\tilde \varphi}_{\alpha}\}$ and two sets of the constraints
are related linearly: ${\tilde \varphi}_{\alpha}= M_{\alpha}^{~\beta}
\varphi_{\beta}$.  If $M_{\alpha}^{~\beta}$ is invertible, two systems
may be considered to be equivalent, since they are defined by the same
sets of the constraint surfaces.\footnote{In the extended phase space
approach to constrained system, the BRST charges of such systems
constructed directly from the first-class constraints are shown to be
related via a canonical transformation acting also on ghost
variables\cite{Henneaux}.}  We establish in this way equivalence
between the D-string and the type IIB Green-Schwarz (GS) string
systems.  In these systems, we have first-class sets of the
constraints corresponding to generators of the reparametrization, the
kappa symmetry, and the U(1) gauge transformation.  There appear also
fermionic second-class constraints in relation to the kappa symmetry.
The separation of the second-class set from the first-class one will
not be discussed in detail, since it is not relevant to the canonical
equivalence.

There are two main results to be reported here. First, the D-string is
equivalent to the type IIB GS superstring. Second, the ``vector duality"
may be elevated to the canonical transformation without resort to any
semi-classical approximations.  The first claim is based on the
observation that two actions have constraints related linearly and thus
define the same constraint surfaces.  The equivalent GS action should
contain a contribution of the world-sheet gauge field, and a simple
candidate for this is the ``theta term". To discuss the second claim on
the ``vector duality", we perform another canonical transformation; the
constraints of the D-string and the mapped constraints of the GS string
share the same dynamical string tension which depends on the electric
field. Once the integer quantization is assumed for the electric field,
the tension is in the $SL(2,Z)$ covariant form\cite{Schwarz} and the
action which reproduces the remaining constraints on the string
variables is the ordinary GS action but with the covariant
tension. However, without the quantization, the action which generates
the full constraints including those on the gauge field is not the GS
type action with the tension replaced by the dynamical one. We give an
action yielding the full constraints: though it takes an unusual form,
it is a supersymmetric extension of the DBI action.

\section{Equivalence between D-string and type IIB string}
\indent

We begin by considering the D-string action with kappa symmetry given
in \cite{Aganagic2} (see also \cite{Witten}). In addition to a
world-sheet gauge field, it contains a dilaton $\phi$ and an axion
$\chi$ backgrounds, which are assumed to be constants.  The action is
given by 
\bea 
S=-n\int d^2\sigma \left\{e^{-\phi}\left[\sqrt{-G_F}+\epsilon^{\mu\nu}
\Omega_{\mu\nu}(\tau_1)\right]+\frac{1}{2}\epsilon^{\mu\nu}
\chi F_{\mu\nu}
\right\} \quad
\label{sdact}
\eea
where
\bea
\begin{array}{rclcrcl}
G_F&=&{\det}(G_{\mu\nu}+{\cal F}_{\mu\nu}) 
   =G+({\cal F}_{01})^2 &,&
G&=&{\det}G_{\mu\nu} =G_{00}G_{11}-(G_{01})^2  \\
G_{\mu\nu}&=&\Pi_{\mu}^m\Pi_{\nu,m}&,&
\Pi_{\mu}^m&=&\partial_{\mu}X^m-\bar{\Theta}\Gamma^m\partial_{\mu}
\Theta
\\
{\cal
 F}_{01}&=&F_{01}-\epsilon^{\mu\nu}\Omega_{\mu\nu}(\tau_3) &,&
F_{\mu\nu}&=&\partial_{\mu} A_{\nu}-\partial_{\nu}A_{\mu}
\end{array}\nonumber \eea 
and 
\bea
\Omega_{\mu\nu}(\tau_1)&=&
\left\{-\frac{1}{2}\bar{\Theta}\Gamma\tau_1\partial_{\mu}{\Theta}
\cdot (\Pi_{\nu}+\frac{1}{2}\bar{\Theta}\Gamma\partial_{\nu}
{\Theta})\right\}-\left\{\mu \leftrightarrow \nu\right\}\nonumber \\
\Omega_{\mu\nu}(\tau_3)&=&\left\{-
\frac{1}{2}\bar{\Theta}\Gamma\tau_3\partial_{\mu}{\Theta} \cdot
(\Pi_{\nu}+\frac{1}{2}\bar{\Theta}\Gamma\partial_{\nu}{\Theta})\right\}
 -\left\{\mu
\leftrightarrow \nu\right\} \quad .  
\eea 
We have used the same conventions
of the Dirac matrices and the $SL(2,R)$ description of N=2 SUSY in
spinor indices as those given in \cite{Aganagic}.

Let $(X^m,~P_m),~(\Theta_{A,\alpha},~\{\mit\Pi_{\Theta}\}_{A,\alpha}),~
(A_{\mu},~E^{\mu})$ be canonically conjugate pairs of the phase
space variables.
The action (\ref{sdact}) is described by a set of constraints 
\bea
{\varphi_0} &=&~\frac{1}{2}[({\cal P})^2+
(T_D)^2 G_{11}] \approx 0 \label{LH}\\
{\varphi_1} &=&~{\cal P}\cdot\Pi_1 \approx 0 \label{LT}\\
{\psi}&=&~{\mit\Pi}_{\Theta}+\bar{\Theta}\Gamma\cdot
({\cal P}-\Pi_1\tau_D) -\frac{1}{2}(
\bar{\Theta}\Gamma{\Theta}'\cdot
\bar{\Theta}\Gamma\tau_D+
\bar{\Theta}\Gamma\tau_D{\Theta}'\cdot
\bar{\Theta}\Gamma )\approx 0\label{F}\\
E^0&\approx&0 \label{Gaussp}\\
\partial_1 E&\approx& 0 \quad ,\label{Gauss}
\eea
where 
\bea
{\cal P}^m=P^m+\bar{\Theta}\Gamma^m\tau_D{\Theta}'
\quad ,\label{cal P}
\eea
$E=E^1$ and ${\Theta}'=\partial_{1}{\Theta}$.
Note that 
in the above set of the
constraints the $\tau$ matrices acting on the $N=2$
spinor indices of $\Theta_A$ 
appear only through the combination of $\tau_D$ defined by
\bea
\tau_D &=&~ (E + n \chi)\tau_3 +n e^{-\phi} \tau_1 \nonumber \\
(\tau_D)^2&=&~(T_D)^2 {\bf 1}_2 \quad, 
\eea
where 
\bea
(T_D)^2~=~(E + n \chi)^2 + (n e^{-\phi})^2 \quad. 
\eea
The string tension of the D-string may be identified with the
co-efficient of $G_{11}$ in the ``Hamiltonian constraint'' (\ref{LH}).
It should be remarked that the D-string has the dynamical tension
$T_D$, which is the desired form in view of the $SL(2,Z)$
covariance.\footnote{The $SL(2,Z)$ covariant tension is $e^{\phi /2}
T_D$.  The difference came from our use of the string metric in the
action; with the Einstein metric in it, one would obtain the covariant 
tension directly in the constraint.}  
The matrix $\tau_D$ is a natural extension of that (defined as $\tau _{E}$)
given in \cite{Hatsuda}.  One requires the constraints (\ref{LH}),
(\ref{LT}), and a half of the fermionic one (\ref{F}) to be first
class, generators of the world-sheet reparametrization and the kappa symmetry
transformation. The remaining half of (\ref{F}) must be second-class
constraints.  This requirement is satisfied if the backgrounds $\phi$
and $\chi$ are constants.  
Canonical quantization of the D-string action
(\ref{sdact}) proceeds in parallel with that discussed in \cite{Hatsuda}.

We establish first canonical equivalence of the D-string action
(\ref{sdact}) with type IIB GS string action with
tension, $T=1$.  It is based on an observation that if one
replaces $\tau_D$ by $\tau_3$ in (\ref{LH}), (\ref{LT}) and
(\ref{F}), one obtains the constraint set of type IIB
superstring. This transformation on the $\tau$ matrix is realized by a
SO(2) rotation around 2-axis \cite{Hatsuda},
\be
U(\rho)=\exp\left(i\tau_2\frac{\rho}{2}\right),
\label{u1-rot}
\ee
with angle fixed by
\be
\cos \rho = \frac{E + n\chi}{T_{D}} \quad, ~~~~~~~~~~~~
\sin \rho = -\frac{n e^{-\phi}}{T_{D}}
\quad .
\label{angle}
\ee
It leads to  
\be
U(\rho)^{-1}~\tau_{D}~U(\rho)= T_{D} \tau_{3}
\quad .
\label{SO(2)rot}
\ee
In order to get the constraint set of the GS string,
 we perform a canonical change of the phase space variables:
\bea
x^m&=&T_{D}^{1/2}~{X}^m  \nonumber \\
p^m&=&T_{D}^{-1/2}~{P}^m   \nonumber \\
\theta&=&T_{D}^{1/4}~U(\rho)^{-1}~{\Theta}\nonumber \\
\mit\Pi_{\theta}&=&T_{D}^{-1/4}~{\mit\Pi_{\Theta}}~U(\rho)  \\
{\hat A}_{0}&=&A_{0}  \nonumber \\
{\hat A}_{1}&=&A_{1}  + \left( \frac{E+n\chi}{2T^2_{D}}\right){P}\cdot
 {X}+{\mit\Pi_{\Theta}}\left[ \left( \frac{E+n\chi}{4T^2_{D}}\right)-
i\tau_{2} \left( \frac{n e^{-\phi}}{2T^2_{D}}\right) \right] {\Theta}  
\nonumber \\
{\hat E}^{\mu}&=&E^{\mu} 
\quad . \nonumber
\label{canotra}
\eea
These are obtained via a generating function
\be
{\hat {\cal W}}=T_{D}^{1/2} p \cdot {X} + {\hat E}^{\mu}A_{\mu} + T_{D}^{1/4} 
\mit\Pi_{\theta} U(\rho)^{-1}~{\Theta}
\quad ,
\label{gene}
\ee
where $T_{D}$ and $U(\rho)$ should be regarded as functions of ${\hat 
E}$ instead of $E$.
When one replaces the D-string variables
$(X,~P,~\Theta,~\mit\Pi_{\Theta})$ by new ones
$(x,~p,~\theta,~\mit\Pi_{\theta})$ in the constraints (\ref{LH}),
(\ref{LT}) and (\ref{F}), there appear terms with derivatives
$\partial_{1} T_{D}$ and $\partial_{1} U(\rho)$. All such terms,
however, are proportional to the gauss law constraint, $\partial_{1} E
=E'$.

It is easy to show that the constraint set of the D-string is 
transformed into that of the GS string supplemented by the gauge field
constraints,
\bea
{\hat \varphi}_{0}&=&\frac{1}{2}[(\wp)^2 + {g}_{11}] \approx 0 
\label{con1} \\
{\hat \varphi}_{1}&=&\wp\cdot {\pi}_{1} \approx 0 \label{con2}\\
{\hat \psi}&=& {\mit\Pi_{\theta}} + \bar{\theta}\Gamma \cdot
(\wp-{\pi}_{1}\tau_3) -\frac{1}{2}(
\bar{\theta}\Gamma{\theta}'\cdot
\bar{\theta}\Gamma\tau_3+
\bar{\theta}\Gamma\tau_3{\theta}'\cdot
\bar{\theta}\Gamma ) \approx 0  \label{fercon}\\
{\hat E}^0 &\approx&0  \label{gauss1} \\
\partial_1 {\hat E}&\approx&0 
\quad ,
\label{gauss2}
\eea
where 
\bea
{\wp}^m=p^m+\bar{\theta}\Gamma^m\tau_3{\theta}'
\qquad {\pi}_{1}^m={x'}^m-\bar{\theta}\Gamma^m\theta'
\qquad g_{11}={\pi}_{1}\cdot{\pi}_{1}
\quad .
\label{wp}
\eea
One obtains the relations
\bea
\varphi_{0}&\sim& T_{D} {\hat \varphi_{0}} \nonumber \\
\varphi_{1}&\sim&  {\hat \varphi}_{1}  \label{contra} \\
\psi&\sim& T_{D}^{1/4}~
{\hat \psi}~ U(\rho)^{-1} \quad , \nonumber \\
\eea
where $\sim$ represents equality up to the gauss law constraint.
Therefore, the new canonical variables,
$(x,~p,~{\theta},~\pi_{\theta})$, turn out to be 
the GS string coordinates and 
their conjugates.
In order to see the connection between the constrained systems of 
the D-string and GS string, we consider 
fermionic constraint algebra under the super-Poisson bracket:
\bea
\{{\psi}_{A,\alpha},{\psi}_{B,\beta}\}&=&
2(C\Xi)_{A,\alpha\ B,\beta} \nonumber \\
\{{\hat \psi}_{A,\alpha},{\hat \psi}_{B,\beta}\}&=&
2(C{\hat \Xi})_{A,\alpha\ B,\beta}
\label{feral}
\eea
where $C$ is the charge conjugation matrix in spinor space, and
\bea
\Xi_{AB}~&=&~\sltP\delta_{AB}-\slPi_1(\tau_D)_{AB} \nonumber \\
{\hat \Xi}_{AB}~&=&~\sltp\delta_{AB}-\slpi_1(\tau_3)_{AB}
\quad .
\label{Xi}
\eea
Note that both matrices are nilpotent, $\Xi^2 = {\hat \Xi}^2 \approx 0$.
 By multiplying $\Xi,~{\hat \Xi}$ to $\psi,~{\hat \psi}$ respectively,
one may project out sets of first-class constraints, $\psi_{I}$
and $\hat \psi_{I}$: 
\bea
\psi_{I}&=& \psi\Xi \nonumber \\
{\hat \psi}_{I}&=& {\hat \psi}{\hat \Xi}=T_{D}^{-1/4}~\psi_{I}~
U(\rho)
\quad .
\label{psifirst}
\eea
It turns out therefore that the bosonic and fermionic first-class
constraints of the D-string are given by a linear combination of those
of the GS string supplemented with U(1) gauge field. The fermionic
second-class constraints of the D-string system are a mixture of the
fermionic first-class and second-class constraints of the GS system.
These linear combinations are clearly invertible.  It follows then
from a general theorem in canonical constrained systems
\cite{Henneaux} that the two systems with vanishing canonical
Hamiltonian define the same constraint surfaces and therefore become
canonically equivalent. The set of fermionic first-class constraints
$\psi_{I}$ in (\ref{psifirst}) is however reducible. Using projection
operator $P_{\pm}=(1\pm \tau_{3})/2$, one obtains an irreducible set
of the first-class constraints, $\psi {\Xi} P_{+}$.  The orthogonal
combination, $\psi_{II}=\psi P_{-}$, gives second-class constraints.
It has been known \cite{Popescu}, \cite{Kallosh}, \cite{Hatsuda} that
this irreducible separation can be done for the D-string system
without spoiling manifest Lorentz covariance. This is because that
$\Xi$ does not commute with $P_{\pm}$. On the other hand, since $\hat
\Xi$ commutes with $P_{\pm}$, one inevitably looses either covariance
or irreducibility in separation of first-class set from second-class
one for the GS string system.

A candidate of GS string action which is canonically 
equivalent to (\ref{sdact}) is
given by
\bea
 S&=&-\int d^2\sigma \left\{\left[\sqrt{-{g}}+\epsilon^{\mu\nu}
 {\omega}_{\mu\nu}(\tau_3)\right] -{\hat E}\epsilon^{\mu\nu}{\hat
F}_{\mu\nu}\right\}
\quad ,
\label{GSaction}
\eea
where ${\omega}_{\mu\nu}(\tau_3)$ is defined in terms of $\theta$, and 
all other variables are defined also using the new canonical variables. 
${\hat
 E}$ is an auxiliary field which becomes the conjugate momentum of
${\hat A}_1$. The ``theta'' term containing ${\hat E}$ is needed to
generate the constraints for the gauge field.
This action is shown to lead to the constraints (\ref{con1}) $\sim$
(\ref{gauss2}). 

\section{Canonical transformation and the S duality}
\indent

In order to discuss $SL(2,Z)$ covariance, we may transform the D-string
described by (\ref{sdact}) into Type IIB string with the same dynamical
 tension $T_{D}$. It is achieved via another canonical
transformation which only rotates the fermionic variables and gauge
field.
\bea
{\check X}^m&=& X^m \nonumber \\ {\check P}^m&=& P^m \nonumber \\
{\vartheta}&=&~U(\rho)^{-1}~{\Theta}\nonumber \\
\mit\Pi_{\vartheta}&=&~\mit\Pi_{\Theta}~U(\rho)  \\
{\check A}_{0}&=& A_{0} \nonumber \\
{\check A}_{1}&=&A_{1}  -i\frac{n e^{-\phi}}{2T_{D}^2}{\mit\Pi_{\Theta}}
\tau_{2} {\Theta}  
\nonumber \\
{\check E}^{\mu}&=&E^{\mu} \nonumber
\quad .
\label{canotra2}
\eea
The generating function of the transformation reads
\be
{\check {\cal W}}= {\check P}\cdot{X} + {\check E}^{\mu}A_{\mu} + 
\mit\Pi_{\vartheta} U(\rho)^{-1}~{\Theta}
\quad .
\label{gene2}
\ee
Using a matrix
\be
\tau_F=\tau_{3}~T_{D}
\quad ,
\label{tauf}
\ee
the constraints are shown to be transformed into
\bea
{\check \varphi}_{0}&=&\frac{1}{2}[({\check{\cal P}})^2 
+ T_{D}^2 {\check G}_{11}] \approx 0 \label{con11} \\
{\check \varphi}_{1}&=& {\check{\cal P}}
\cdot {\check \Pi}_{1}\approx 0 \label{con22}\\
{\check \psi}&=& {\mit\Pi_{\vartheta}} + \bar{\vartheta}\Gamma \cdot
({\check{\cal P}}-{\check \Pi}_1\tau_F)
-\frac{1}{2}(
\bar{\vartheta}\Gamma{\vartheta}'\cdot
\bar{\vartheta}\Gamma\tau_F+
\bar{\vartheta}\Gamma\tau_F{\vartheta}'\cdot
\bar{\vartheta}\Gamma )\approx 0 \label{fercon2}\\
{\check E}^0&\approx&0 \label{gauss12} \\
\partial_1 {\check E}&\approx&0 
\quad ,
\label{gauss22}
\eea
where 
\bea
{\check{\cal P}}^m=P^m+\bar{\vartheta}\Gamma^m\tau_F{\vartheta}'
\qquad 
{\check \Pi}_{1}^m={X'}^m-\bar{\vartheta}\Gamma^m{\vartheta}'
\qquad 
{\check G}_{11}={\check \Pi}_{1}\cdot{\check \Pi}_{1}
\quad .
\label{checkP}
\eea
The constraints (\ref{con11}) $\sim$ (\ref{gauss22}) are specified by
 $\tau_F$, while those for the D-string (\ref{LH}) $\sim$
 (\ref{Gauss}) by $\tau_D$. Although they give rise to the same
 dynamical tension ${\tau_{D}}^2 = {\tau_{F}}^2 = T_{D}^2~{\bf
 1}_{2}$, they differ in the matrix structure which corresponds
 to the SO(2) rotation of the fermionic variables.

An action which generates the above set of the constraints reads
\bea
 S&=&-n\int d^2\sigma \left\{e^{-\phi}
\sqrt{\left
(\sqrt{-\check G}+\epsilon^{\mu\nu}
{\check \Omega}_{\mu\nu}
(\tau_{3})\right)^2 -\left(\frac{1}{2}\epsilon^{\mu\nu}{\check
F}_{\mu\nu}\right)^2}
+\frac{1}{2}\chi\epsilon^{\mu\nu}{\check
F}_{\mu\nu}\right\}
\quad.
\label{DGSaction}
\eea
It should be remarked that the bosonic truncation reduces this action
to the conventional DBI action. Therefore, (\ref{DGSaction}) gives a
supersymmetric extension of the DBI action. 

Comparing (\ref{con11}) $\sim$ (\ref{gauss22}) with (\ref{con1}) 
$\sim$ (\ref{gauss1}), one recognizes that the above constrained
system describes the GS string with dynamical tension $T_{D}$.
 The tension becomes the $SL(2,Z)$ covariant
form, if the electric field is
quantized to be integer values.  
 Quantization of U(1) gauge field can
be done, compactifying the world-sheet co-ordinate $\sigma$ into a
circle (of length $2\pi$).  One takes the temporary gauge, ${\check
A}_{0}=0$, leaving ${\check A}_{1}$ subjected to the periodic boundary
condition, ${\check A}_{1}(\tau, \sigma)={\check A}_{1}(\tau, \sigma +
2\pi) $.  A large gauge transformation, ${\check A}_{1} \rightarrow
{\check A}_{1} + m$ with integer $m$, is known to only acts as a
constant shift for the zero mode,
\be
{\check {\cal A}}_{1} =  \int d \sigma {\check A}_{1} 
\rightarrow {\check {\cal A}}_{1} + 2\pi m
\quad .
\label{zeromode}
\ee
 If one identifies the zero mode configurations connected mutually via
the large gauge transformations, ${\check {\cal A}}_{1}$ becomes an
angular variable. The conjugate zero mode momentum ${\check {\cal
E}}^{1}$, therefore, is quantized to be an integer value $m$. The
Hamiltonian of the system is a function of the electric field ${\check
E}^{1}$, and diagonalized in the momentum representation. Clearly,
only the ${\check {\cal E}}^{1}$ dependence can survive in the
physical state through the gauss-law constraint,
\be
\partial_{1}{\check E}^{1} |~phys> =0
\quad .
\label{gaussconstraint}    
\ee
This gives the integer quantization of the electric field, generating
NS-NS charge, ${\check E}^{1}=m$. 

Although the constrained system with (\ref{con11}) $\sim$
(\ref{gauss22}) describes manifestly the $SL(2,Z)$ covariance of the
D-string theory, one can not immediately see the covariance in the
action (\ref{DGSaction}). If one performs, however, the electric field
quantization described above first, one is left with the constrained
system corresponding to the action:
\bea
 S&=&-\int d^2\sigma \left\{\sqrt{(m+n\chi)^2 + (ne^{-\phi})^2}
\left[\sqrt{-{\check G}}+\epsilon^{\mu\nu}
{\check \Omega}_{\mu\nu}(\tau_3)\right]\right\}
\quad .
\label{RGSaction}
\eea
We emphasize that this form of action may be written only after the
canonical transformation; the crucial element is the SO(2) rotation
(\ref{canotra2}) which changes the $\tau$-matrix structure from
$\tau_D$ to $\tau_F$.
 
\section{Discussion}
\indent

We have constructed two sets of the canonical transformations: the first
one directly exhibits the equivalence of the D-string with the type IIB
string with the normal tension;
the second one is given to
replace the vector-duality transformation.  
In both cases, the
transformations in fermionic sector contain an SO(2) rotation
around 2 axis of N=2 spinor space. The approach given here is not
based on any semi-classical approximations, and the results precisely
agree with the $SL(2,Z)$ covariance of D-string. We close this paper with
the following remarks.
\begin{enumerate}
\item
If one wishes to show the equivalence between two theories more
explicitely, one may  
employ BRST quantization scheme. After performing the canonical
transformation between the string variables, one eliminates the
fermionic second-class constraints, leaving first-class set of the
constraints for each system.  One is given by a linear combination of
the other.  Since this relation is clearly invertible as long as $E
\ne 0$, two BRST charges constructed from different sets of the
constraints are related by a canonical transformation defined in
extended phase space, as a result of the general theorem given in
\cite{Henneaux}. Construction of the BRST charges is tedious but important
to compare two theories.\footnote{The on-shell nilpotent BRST charge
for the super D-string is given in \cite{Hatsuda}.}  In suitable gauge
choices, one may discuss quantization of the D-string system and the
GS string system, and examine their spectrum as well as the structure
of the global supersymmetry. A crucial difference between two systems
is that the D-string system can be covariantly quantized with finite
ghost variables, unlike the GS system.
\item 
We have been restricted ourselves to the flat target-space metric
and constant backgrounds of the dilaton and axion fields.  In
addition, the other backgrounds such as anti-symmetric tensor fields
have been ignored here. These restrictions should be removed to
establish the S duality.
\item
There have been some attempts to construct manifestly $SL(2,Z)$
covariant superstrings where additional world-sheet gauge field is
introduced to respect the $SL(2,Z)$ symmetry more
explicitely \cite{Townsend}. It will be possible to give an extended
canonical transformation, which connects such actions with some
extension of the GS action with a couple of world-sheet gauge
fields.
\item
It is important to examine if our canonical transformation
approach given here for the D-string applies to the D3-brane and other brane
actions in type IIB theory. As will be reported in a subsequent
publication \cite{Igarashi}, canonical transformation can be
construnted to show the $SL(2,Z)$ invarinace of the D3-brane action.
For even p-branes such as D2-brane and D4-brane, investigations of the
relationship between the type IIA theory and a compactified $D=11$ M
theory based on canonical transformation will be especially
interesting.
\end{enumerate}

\medskip\noindent
{\bf Acknowledgments}\par
\medskip\par

K. I. would like to thank to YITP for its kind hospitality extended to him.
R. K. is obliged to A.Sugamoto for enjoyable conversations
on dualities in string theories and for encouragement.

\end{document}